\documentclass[aps,prd,showpacs,preprintnumbers,nofootinbib]{revtex4}
\pdfoutput=1
\usepackage{graphicx}
\usepackage{bm,latexsym,amsmath,amssymb,amsfonts}
\usepackage{subfig}
\usepackage[colorlinks]{hyperref}
\usepackage{color}

\def\prd{Phys.~Rev.~D}
\def\prl{Phys.~Rev.~Lett.}

\def\cqg{Class.~Quant.~Grav.}
\def\lrl{Living.~Rev.~Relativity.}
\def\ijmpd{Int. ~J. ~Mod. ~Phys. ~D.}

\begin{document}
\preprint{KUNS2555, YITP-15-29}

\title{Synergy between ground and space based gravitational wave detectors 
	for estimation of binary coalescence parameters}

\author{ Remya Nair$^{1}$, Sanjay Jhingan$^{1}$, and Takahiro Tanaka$^{2,3}$ }
\affiliation{$^{1}$Centre for Theoretical Physics, Jamia Millia Islamia, New Delhi -25, India }
\affiliation{$^{2}$Yukawa Institute for Theoretical Physics, Kyoto University, 606-8502, Kyoto, Japan }
\affiliation{$^{3}$Department of Physics, Kyoto University, 606-8502, Kyoto, Japan}

\begin{abstract}
We study the advantage of the co-existence of future ground and space based 
gravitational wave detectors, in estimating the parameters of a binary 
coalescence. Using the post-Newtonian waveform for the inspiral 
of non-spinning neutron star-black hole  pairs in circular orbits, we study 
how the estimates for chirp mass, symmetric mass ratio, and time and phase at coalescence
are improved by combining the data from different space-ground detector pairs. 
Since the gravitational waves produced by binary coalescence also provide a suitable domain 
where we can study strong field gravity, we also study the deviations from general relativity using the parameterized post-Einsteinian framework. As an example, focusing on the 
Einstein telescope and DECIGO pair, we demonstrate that there exists a sweet spot range of sensitivity in the pre-DECIGO phase where the best enhancement due to the synergy effect 
can be obtained for the estimates of the post-Newtonian waveform parameters as well as the modification parameters to general relativity.
\end{abstract}
\maketitle

\section{Introduction}
General relativity (GR) has enjoyed a very successful run as a theory 
describing the classical Universe by passing all the tests 
available so far \cite{will}. 
Yet two of the unique predictions of GR, the black 
holes (BHs) and gravitational waves (GWs), are waiting for direct confirmation. 
There is an ever increasing indirect evidence of stellar mass BHs 
as well as 
super massive BHs at the center of each galaxy \cite{RN}. Similarly, the Universe 
is immersed in a GW background, analogous to the cosmic microwave 
background, but we still have no glimpse of it. 
Thanks to several decades of 
research developing detectors with ever increasing sensitivity, 
we are likely to be very close to detecting GWs and 
to open a new era in astronomy. 
It will not be an overstatement to say that if the 20th century belonged
to astronomy with 
electromagnetic waves, the 21st century would definitely belong to astronomy 
with GWs \cite{kent,will_yunes}. The detection of GWs would not only be a triumph of GR as the theory 
describing our classical world, but it would also give us a unique opportunity 
to detect BHs by directly mapping their spacetime.

GR has been subjected to a plethora of experimental tests, but mostly
the explored regimes are those in which the gravitational field is weak 
and particle velocities are small relative to the speed of light \cite{will}.
Thus, one of the most promising outcomes of GW astronomy would be the
possibility of testing the validity of GR in the dynamical strong field
regime \cite{siemens}. The prospect of discovering that GR may need some modification
in the strong field regime has also motivated the development of several
alternative theories of gravity. 
GW signals will provide a testing ground for these theories and
a wide range of tests of GR have already been proposed. Tests that use  GWs coming from compact binary coalescence include those proposed in \cite{grt1,grt2,grt3,grt4,ppn2,ppn1,YT}. The coalescing 
binaries composed of neutron stars (NS) and BHs are amongst the
most likely sources for GW signals to be observed by ground-based 
interferometric detectors. When observed, these GW signals 
will probe the strong field regime of gravity and allow the 
implementation of these tests of GR.

The post-Newtonian formalism (PN) has been used to model the 
inspiral part of the binary evolution within GR. 
In the PN formalism physical quantities of interest such 
as the conserved energy, flux etc. are found
as expansions in the small parameter $(v/c)$, where $v$ is the 
characteristic speed of the binary 
system and $c$ is the speed of light \cite{luc}. 
In standard convention $O((v/c)^n)$ corrections counting 
from the leading order are referred to as a 
$(n/2)$PN order term. The expansion of the GW waveform is 
currently known up to 3.5PN, and the binary motion is known up to 
4PN order \cite{luc}.
Inspiraling binaries are suitable for studying the strong field regime. 
The orbital velocities in these systems can go as high as $0.5c$, 
and therefore higher order PN corrections are relevant. 
The standard data analysis techniques 
used for the detection and analysis of GW signals depend 
on the availability of accurate templates to identify weak signals buried in the noise. 
However, in order to carry out the tests of GR,  
it is impractical to make accurate templates for all possible
alternative theories of GR. A more feasible way of 
carrying such tests is to adopt general non-GR templates to model the 
signals. Arun et al., proposed such templates in which 
the expansion coefficients of the phase in the Fourier transform of the
GW waveform (see equation (\ref{phase}) ahead)
are treated as fitting parameters~\cite{grt4}. In GR each 
coefficient is specified by the masses and spins of the constituting 
compact objects of the binary. 
In principle, this relation can be completely different in 
some alternative theory and may even involve other parameters. 

Yunes and Pretorius developed the parameterized post-Einsteinian (ppE) framework, 
which allows for characterizing a wider range of deviations to the
amplitude and phase of the waveform \cite{ppn2}. 
Similar to the parameterized post-Newtonian framework 
they introduced ppE parameters, but instead of parametrizing the metric
tensor, they parametrize the GW response function using a 
generic template family (see section \ref{secppe} ahead). This family can accommodate 
the inspiral phase of most of the known alternative theories with appropriate choices of 
the parameters. 
Cornish et al. applied the ppE approach to simulated data to determine the level 
at which departures from GR can be detected, and also analysed the bias introduced 
in the extraction of the model parameters due to the assumption of the wrong 
theory \cite{ppn1}.

In this work we study the synergy between the ground and space based GW detectors 
in estimating the parameters of the inspiraling binaries, to see whether there is 
any gain in combining measurements from two detectors. 
For this we use the ppE framework developed by
Yunes and Pretorius \cite{ppn2}. We consider two future space based detectors (DECIGO and eLISA) and two future 
ground based detectors (Advanced LIGO and ET). As representative test cases we 
consider two binary systems, a $1.4~M_{\odot}+10~M_{\odot}$ binary and a 
$1.4~M_{\odot}+100~M_{\odot}$ binary, where $M_{\odot}$ is the solar mass. Within the 
PN framework, we study the synergy effect in detail for the chirp mass, the symmetric 
mass ratio and the time and phase at coalescence, and within the
ppE framework we study the synergy effect on 
the estimation of the ppE parameter characterizing deviation from GR. 
The plan of the paper is as follows: in the next section we briefly 
describe the waveforms used in this analysis (\ref{secpn}, \ref{secppe}), the noise curves 
for the different detectors (\ref{secnoise}), and the methodology for error 
estimation (\ref{secdata}). We summarise our results in section \ref{result}.

\section{Methodology}
\subsection{PN formalism}\label{secpn}
We use the PN expression for the Fourier transform $H(f)$, of the
GW signal coming from the inspiral of binary compact objects in
a circular orbit (under the stationary phase approximation)~\cite{satya}. 
Moreover, the \emph{restricted PN waveforms} are used which keep the
higher terms in phase but only take the leading terms for the amplitude
\cite{cutler}. This is because 
in the matched filtering analysis used for the parameter estimation of
binaries the correlation of two waveforms is more sensitive to
deviation in phase than the amplitude. This is not true for
binaries with misaligned spins, where the amplitude modulation on
precession time scale is  also important to determine the spin
parameters \cite{veccio}. As we consider non-spinning binaries for simplicity, 
this modulation is beyond the scope of this paper.  

Hence we consider only the leading order term in the amplitude, while the phase 
terms are taken up to 3.5PN order (note that terms with exponent $k=n$
are ($n/2$) PN order terms).
Namely, we adopt the waveform 
\begin{equation}
H(f)= {\cal A} f^{-7/6} \exp(i \psi (f) + i \pi/4),
\label{waveform}
\end{equation}
where the Fourier amplitude $\cal{A}$ is given by 
\begin{equation}
{\cal A}=\frac{{\cal C}}{D\pi^{2/3}}\sqrt{\frac{5 \nu}{24}} M^{5/6},
\label{amp}
\end{equation}
and the phase $\psi(f)$ is given by
\begin{equation}
\psi(f)=2 \pi f t_c + \phi_ c + \frac{3}{128 \nu} \sum\limits_{k=0}^7 
\alpha_k (\pi M f)^{(k-5)/3}.
\label{phase}
\end{equation}
Here ${\cal C}$ is an $O(1)$ dimensionless geometric factor that depends
on the relative orientation of the binary and the
detector (average over all orientations $\bar{\cal C} = 2/5$), $\nu$ is the
symmetric mass ratio, $M$ is the total mass of the binary, $D$ is the
luminosity distance and $t_c$ and $\phi_c$ are the time and phase at
coalescence, respectively. For the purpose of this paper, 
the explicit expression for ${\cal C}$ is not necessary as we also neglect
the effects of the orbital motion of the space antenna. 
The expressions for the amplitude and the phase are often also expressed 
in terms of the chirp mass ${\cal M}=\nu^{3/5}M$ as
\begin{equation}
{\cal A}=\frac{C}{D\pi^{2/3}}\sqrt{\frac{5}{24}} {\cal M}^{5/6},
\label{amp2}
\end{equation}
and 
\begin{equation}
\psi(f)=2 \pi f t_c + \phi_ c + \frac{3}{128 \nu} \sum\limits_{k=0}^7 
\alpha_k \left(\frac{\pi {\cal M} f}{\nu^{3/5}}\right)^{(k-5)/3}.
\label{phase2}
\end{equation}
The values of the coefficients in the PN expansion of the Fourier phase are 
as follows \cite{satya}:
\begin{eqnarray}
\alpha _0 &=& 1, ~~\alpha _1=0, ~~\alpha _2 = \frac{3715}{756}+
\frac{55}{9}\nu ,\nonumber \\
\alpha _3 &=& -16 \pi ,
~~\alpha _4 = \frac{15293365}{508032}+\frac{27145}{504}\nu + 
\frac{3085}{72}\nu ^2, \nonumber \\
\alpha  _5 &=& \pi \left(\frac{38645}{756}-\frac{65}{9}\nu\right)
\left[1+\ln \left(6^{3/2}\pi M f\right) \right], \nonumber \\
\alpha _6 &=& \frac{11583231236531}{4694215680}-\frac{640}{3} \pi^2 -
\frac{6848}{21}\gamma + \left(-\frac{15737765635}{3048192}+
\frac{2255}{12}\pi^2 \right)\nu \nonumber \\
&&+ \frac{76055}{1728}\nu^2 -\frac{127825}{1296}\nu^3 - 
\frac{6848}{63}\ln (64\pi M f), \nonumber \\
\alpha _7 &=& \pi \left( \frac{77096675}{254016}+
\frac{378515}{1512}\nu - \frac{74045}{756}\nu^2\right).
\end{eqnarray}
As mentioned earlier, we consider two binary systems, a $1.4 ~M_{\odot}+10 
~M_{\odot}$ NS-BH binary and a $1.4 ~M_{\odot}+100~M_{\odot}$ NS-BH binary. 

\subsection{Corrections due to modified gravity: ppE formalism}\label{secppe}
One of the most important outcomes of GW detection would be placing bounds 
on alternative theories of gravity~\cite{NordWill}. Several such
theories have been proposed to explain the observed late time
acceleration of the Universe. Within the framework of GR the explanation
is given by assuming exotic matter fields, but if classical GR is
realised as a low energy limit of some more 
fundamental theory, we might expect non-trivial modifications to it 
({\em e.g}. non-minimally coupled dilatons etc.).  

GR has been tested with high accuracy in solar system and also 
through binary pulsars, and in both cases 
gravitational fields are weak 
and particles velocities are small~\cite{will}. 
The lack of experimental verification of GR in strong fields may not
be a problem for long once we start to observe GWs directly. 
In particular, GW detectors in space like eLISA and DECIGO 
will be able to map the spacetime geometry of super-massive BHs at the center of galaxies. This can be done  
very accurately by observing GWs emitted by their inspiraling 
satellites, allowing us to test alternative theories of gravity. 

For studying the modification to GR using a parameterized approach we 
follow the ppE formalism. The modification to the 
Fourier transform $H(f)$ of the GW signal (under the stationary 
phase approximation) from the inspiraling binary~\eqref{waveform}
would be characterized by the leading order corrections 
to the Fourier amplitude and the phase as
\begin{equation}
{\cal A}=\left(1+ \sum_i \alpha_i (\pi {\cal M} f)^{a_i}\right)A_{GR}(f),
\end{equation}
and
\begin{equation}
\psi(f)=\left(\psi_{GR}(f)+\sum_i \beta_i (\pi {\cal M} f)^{b_i}\right).
\label{ppe}
\end{equation}
Here $A_{GR}(f)$ 
is given by equation (\ref{amp}), and 
$\psi_{GR}(f)$ is given by equation (\ref{phase}), the standard GR 
expressions. The coefficients $\alpha_i$ and $\beta_i$, in general, may 
depend on parameters such as chirp mass, spin angular momentum etc. As 
explained in previous subsection, for the present analysis we concentrate 
on the modification in the phase only, meaning $A(f)=A_{GR}(f)$, i.e., 
$\alpha_i =0$.  
\subsection{Noise curves}\label{secnoise}
The output of a GW detector $s(t)$, can be decomposed into two components:
the GW signal $h(t)$, and the detector noise $n(t)$, i.e. $s(t)=h(t)+n(t)$. 
For simplicity, we assume that the detector noise is stationary and Gaussian. 
Stationarity implies that the different Fourier components of the noise are 
uncorrelated, and hence the one-sided noise power spectral density (PSD)
$S_n(f)$ can be defined by
\begin{equation*}
\left\langle \tilde{n}(f)\tilde{n}(f') \right\rangle =
\frac{1}{2}\delta(f-f')S_n(f),
\end{equation*}
where $\tilde{n}(f)$ is the Fourier transform of the noise $n(t)$. 
A commonly used quantity to describe sensitivity curve 
is square root of PSD in units of Hz$^{-1/2}$, and when integrated over positive 
frequencies it gives mean square amplitude of the signal in detector~\cite{moore}. 
We describe the noise spectral densities 
of all the detectors used in this study below. 

\subsubsection*{{\bf Future space based detectors}}
{\bf DECIGO}: The Deci-hertz Interferometer Gravitational Wave
Observatory (DECIGO) is a future plan of a space
mission for observing GWs around $f \sim 0.1-10 Hz$, proposed
by~\cite{Seto:2001qf}. The following noise curve is taken from
\cite{seto}
\begin{equation}
S_n(f)=6.53 \times 10^{-49} \left[ 1+ \left(\frac{f}{f_p}\right)^2 \right]
+ 4.45 \times 10^{-51} \left(\frac{f}{1 \mbox{Hz}}\right)^{-4} \frac{1}
{1+\left(\frac{f}{f_p}\right)^2}+4.94 \times 10^{-52}\left(\frac{f}{1 \mbox{Hz}}\right)^{-4} 
\mbox{Hz}^{-1},
\end{equation}
where $f_p = 7.36$Hz.\\

{\bf eLISA}: 
eLISA is a slightly down-graded mission which is
derived from the Laser Interferometer Space Antenna (LISA) proposal. 
This mission is targeted at the low-frequency band: $0.1$mHz to $1$Hz. 
The following noise curve is taken from \cite{elisa}:\\
\begin{equation}
S_n(f)= \frac{20}{4} \frac{4 S_{acc}(f)/(2\pi f)^4+S_{sn}(f)+S_{omn}(f)}
{L^2} \left( 1+ \frac{f}{0.41 c/2L}\right)^2,
\end{equation}
with 
\begin{align}
& S_{acc}(f)=2.13 \times 10^{-29}(1+10^{-4}\mbox{Hz}/f) \mbox{m}^2 \mbox{s}^{-4} \mbox{Hz}^{-1},\cr
& S_{sn}(f)=5.25 \times 10^{-23}\mbox{m}^2 \mbox{Hz}^{-1},\cr
& S_{omn}(f)=6.28 \times 10^{-23}\mbox{m}^2 \mbox{Hz}^{-1}.
\end{align}
where $L=10^6$km and $c$ is the speed of light.\\

\subsubsection*{{\bf Future ground based detectors}}
{\bf Advanced LIGO}: 
As the representative of the second generation of ground-based
Laser Interferometer Gravitational-Wave Observatory (LIGO) 
The noise curve for the advanced LIGO detectors is taken from \cite{ajith}.\\
\begin{equation}
S_n(f)=10^{-49} \left[\left(\frac{f}{f_0}\right)^{-4.14} -5 \left(
\frac{f}{f_0}\right)^{-2} +111 \left(\frac{1-\left(\frac{f}{f_0}\right)^{2}+
\left(\frac{f}{f_0}\right)^{4}/2 }{1+\left(\frac{f}{f_0}\right)^{2}/2}\right) \right],
\end{equation}
where $f_0=215$Hz.\\

\begin{figure}[h]
\centering  
{\includegraphics[width=3.4in,height=2.4in]{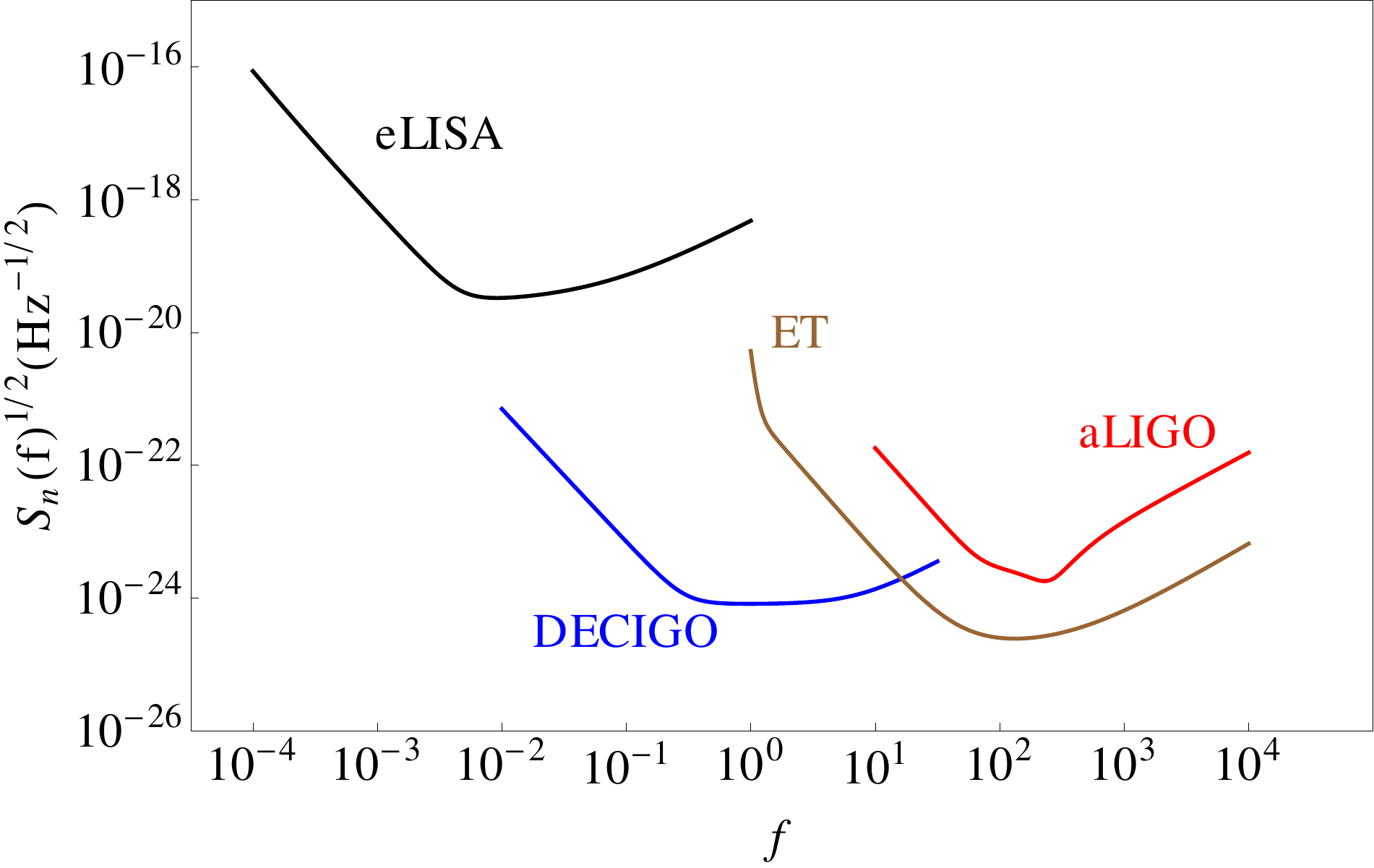}}
\caption{Noise spectra for the different GW detectors used in this work.}
\label{nc}
\end{figure}

{\bf ET}: The Einstein Telescope (ET) is a European- commission project which aims at 
developing the third generation Gravitational Wave observatory. The following noise curve
is taken from \cite{ajith}:\\
\begin{equation}
S_n(f)=10^{-50} \left[2.39 \times 10^{-27}\left(\frac{f}{f_0}\right)^{-15.64} +0.349 
\left(\frac{f}{f_0}\right)^{-2.145} +
1.76 \left(\frac{f}{f_0}\right)^{-0.12} +0.409 \left(\frac{f}{f_0}\right)^{1.1} 
\right]^2,
\end{equation}
where $f_0=100$Hz. The noise curves for the various detectors are plotted in figure (\ref{nc}).

\subsection{Likelihood function and Fisher matrix}\label{secdata}
In this section we briefly overview parameter estimation and Fisher matrix method of error 
estimation. For a review on this topic, we refer to \cite{finn}. 
We assume that the GW signal depends on the parameter vector $\boldsymbol{\theta}$.
In case of the PN waveform $\boldsymbol{\theta}=\lbrace {\cal M}, \nu,
t_c, \phi_c \rbrace$, 
and in case of the ppE framework $\boldsymbol{\theta}=\lbrace {\cal M}, \nu, t_c, 
\phi_c, \beta_i \rbrace$ 
(in this work the ppE parameter $\alpha_i$ is fixed 
to zero for all $i$). Now the noise weighted inner product
for two signals $h_1(t)$ and $h_2(t)$ is defined as:
\begin{equation}
(h_1|h_2)=2 \int _0 ^{\infty} \frac{\tilde{h}_1^*(f) \tilde{h}_2(f)+
\tilde{h}_2^*(f) \tilde{h}_1(f)}{S_n(f)}df,
\end{equation}
where $\tilde{h}_1(f)$ and $\tilde{h}_2(f)$ are the Fourier transforms of
$h_1(t)$ and $h_2(t)$, respectively, and $^*$ represents 
the complex conjugation. The inner product is defined so that the
probability that the signal depends on the parameters $\boldsymbol{\theta}$
(assuming the noise is Gaussian) is given by
\begin{equation*}
P(s|\boldsymbol{\theta})\propto e^{-(s-h(\boldsymbol{\theta}) | 
s-h(\boldsymbol{\theta}))/2}.
\end{equation*}
Hence, the inner product is similar to defining a chi-squared merit
function. 
Given an output $s(t)$, the best fit GW waveform $h(\boldsymbol{\theta})$ 
is obtained by minimizing this inner product (\cite{finn}). For
different realizations
of noise, we may obtain slightly different values of the parameters, but for
large signal to noise ratio they will all be centred around the correct values 
say $\bar{\boldsymbol{\theta}}$ with some spread $\Delta\boldsymbol{\theta}$.
The estimation errors follow a Gaussian distribution 
\begin{equation}
P(\Delta\theta^i)\propto e^{-\Gamma_{ij}\Delta\theta^i\Delta\theta^j /2},
\end{equation}
where $\Gamma_{ij}$ is the Fisher matrix defined as (\cite{cutler})
\begin{equation}
\Gamma_{ij}\equiv\left(\frac{\partial h}{\partial \theta_i}\bigg|
\frac{\partial h}{\partial \theta_j}\right).
\end{equation}
The covariance matrix $C$ is the inverse of the Fisher matrix, i.e.  $C=\Gamma^{-1}$,
and the root mean square error of the parameters can be evaluated from the diagonal elements as
\begin{equation}
\sqrt{\left\langle(\Delta\theta^i)^2\right\rangle}=\sqrt{C^{ii}}.
\end{equation}
When analysing the synergy effect of combining the  measurements from 
two detectors, we take the sum of the Fisher matrices corresponding to 
them and then take the inverse of the matrix thus obtained:
\begin{eqnarray}
\Gamma_{\rm comb} &= \Gamma_{1}+\Gamma_{2},\\
C_{\rm comb} &= \Gamma_{\rm comb}^{-1}.
\label{Ccomb}
\end{eqnarray}
\subsection*{Studying the change in estimates by varying design 
sensitivity of DECIGO}
We also study the synergy effect on the error estimate of various parameters 
by varying the sensitivities of the future space mission DECIGO (and its precursor
pre-DECIGO), since its design sensitivity
is still to be determined depending on the scientific gain. 
Here, for brevity, we simply call it DECIGO. 
In both better and worse directions, the sensitivities are changed 
by scaling the fiducial noise curve uniformly 
over all frequencies. Namely, the scaled DECIGO noise curves are obtained as
\begin{equation}
S_n(f)^{scaled}={\cal K} S_n(f)^{DECIGO}, 
\label{sdecigo}
\end{equation}
with a constant ${\cal K}$. 

\section{Results}\label{result}
In this section we explore the expected errors in the parameter estimation 
when we combine outputs of ground and space based detectors. 
We discuss results on the various parameters in
the PN and the ppE waveforms for coalescing binary systems.  
As mentioned earlier we consider two NS-BH binary systems: a 
$1.4~M_{solar}+10~M_{solar}$ binary and a $1.4~M_{solar}+100~M_{solar}$
binary with the distance fixed at $200$ Mpc. We fix the lower frequency limits at $0.1$ Hz for the space based GW detectors. The space based detectors are expected to be sensitive at
frequencies much below this cut off (eLISA is expected to be sensitive at
as low as $0.1$ mHz), but we chose this cut off since the evolution of the binary coalescence is really slow at lower frequencies. For the mass configurations we have considered in this work, it will take many years
for the binaries to span the lower ($10^{-5}$ - $0.1$) frequency 
range and we would not be able to observe this continuously. 
The evolution is much faster in the mid-high frequency range and
there is a chance we can observe the same events in both space and
ground based detectors. For advanced-LIGO we set the lower frequency limit at $10$ Hz.
\subsection{Parameter estimation - PN expansion}
Here we present the error estimates on 
the chirp mass, the symmetric mass ratio, and the phase and time at coalescence 
using the method described in the previous section.
In table \ref{tab1} and \ref{tab2} we tabulate these errors for both
independent and combined measurements.
We note that there is some gain in combining space-ground measurements for all parameters. In case of the chirp mass, maximum gain is obtained for the detector pair Advanced-LIGO - eLISA, while for all the other pairs one of the GW detector completely dominates the error estimate. Similarly for the time of coalescence maximum gain is obtained for the detector pair ET-DECIGO, and for the other pairs the estimates are dominated by one of the two detectors. For the symmetric mass ratio and the phase of coalescence there is no substantial gain in combining the space-ground measurements since one of the detector always dominates the results. 
\begin{figure}[h]
\centering \subfloat[Part 3][]{
\includegraphics[width=3.5in,height=2.5in]{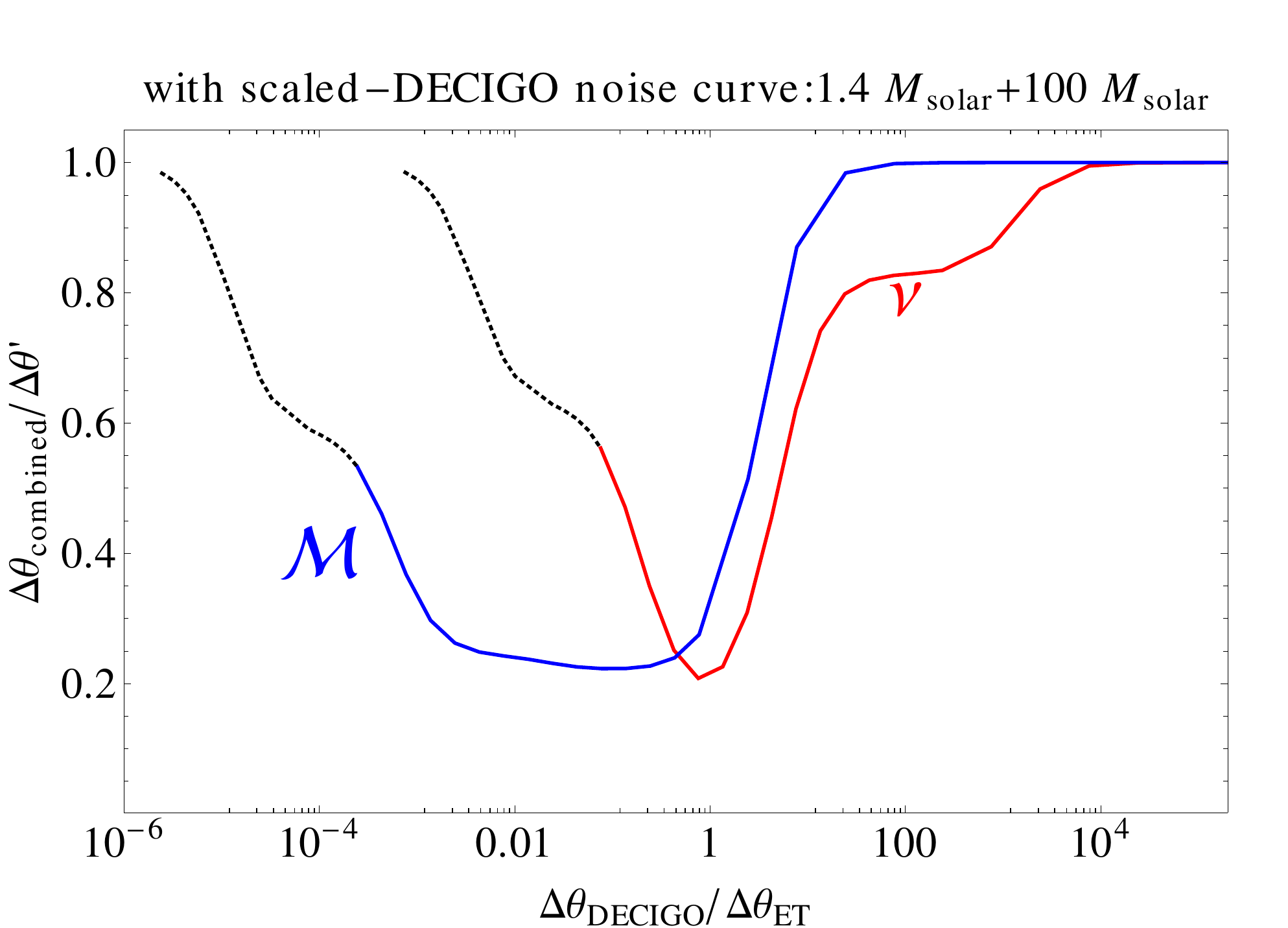}} ~~~~~~~~~~~~~
\subfloat[Part3][]{\includegraphics[width=3.5in,height=2.5in]{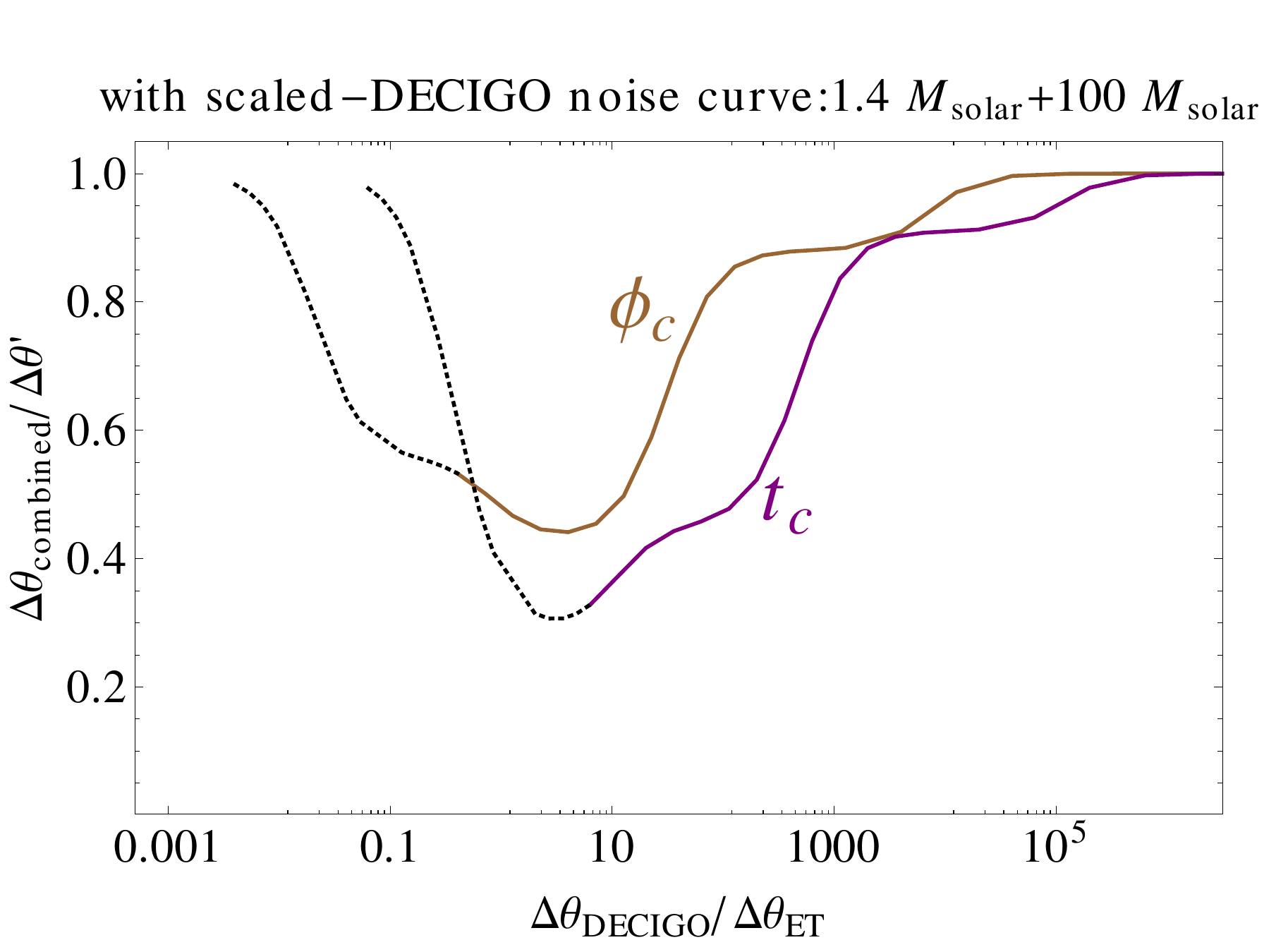}}
\caption{Variation in the error estimates on the chirp mass and the 
symmetric mass ratio (a), and the time and phase at coalescence (b), obtained 
by varying DECIGO sensitivities. $\Delta \theta_{\rm comb}$
($\theta={\cal M}/\nu/t_c/p_c$) is obtained as explained in section \ref{secdata}, 
and $\Delta \theta '$ is as given in equation (\ref{thetap}). The ratio
$\Delta \theta_{\rm comb}/\Delta \theta '$ is plotted against $\Delta \theta$ 
as obtained from the {\it scaled} DECIGO measurements. The dotted curves 
correspond to better sensitivity than DECIGO}
\label{figmcnu2}
\end{figure}
\begin{figure}[h]
\centering \subfloat[Part 3][]{
\includegraphics[width=3.5in,height=2.5in]{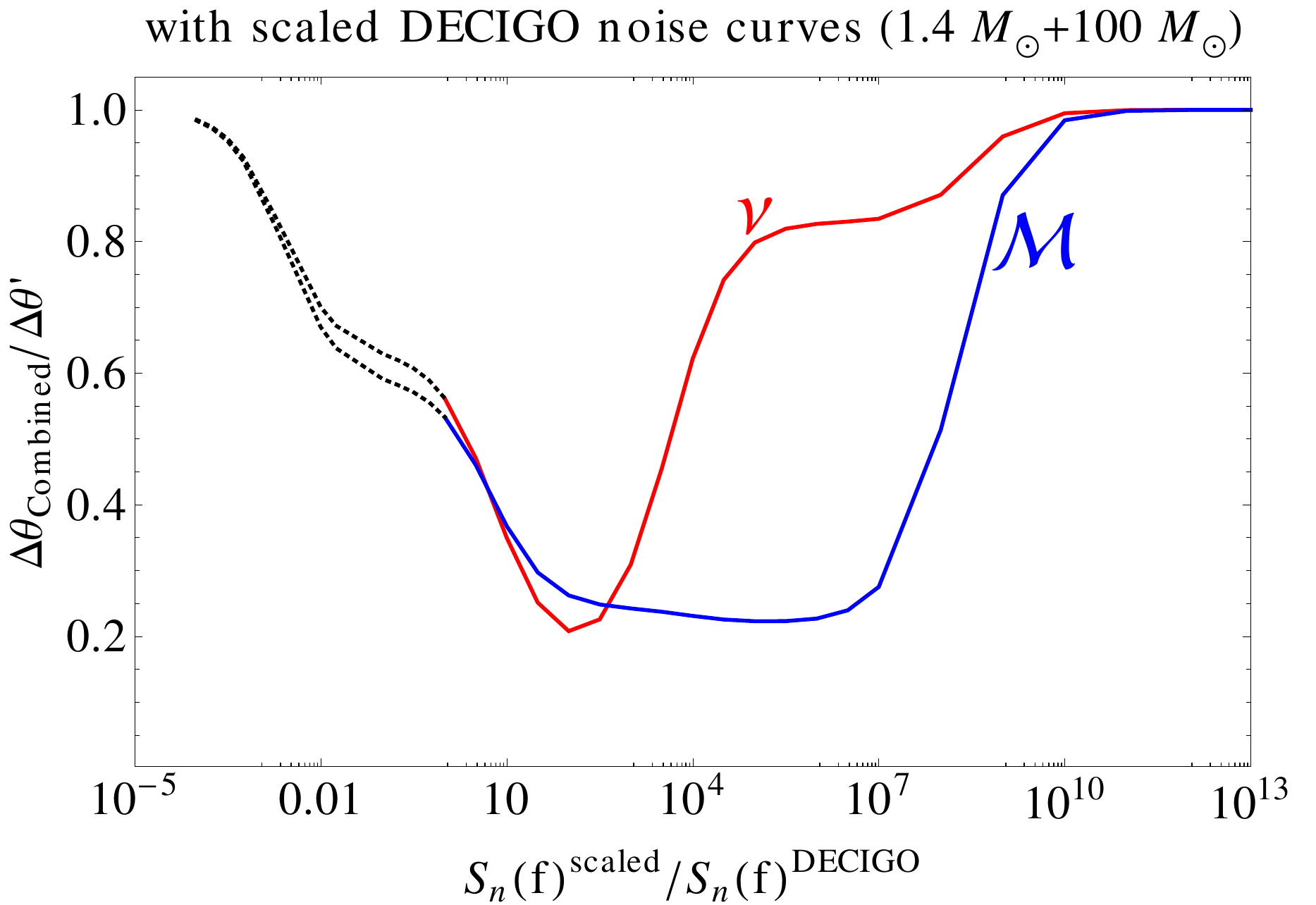}} ~~~~~~~~~~~~~
\subfloat[Part3][]{\includegraphics[width=3.5in,height=2.5in]{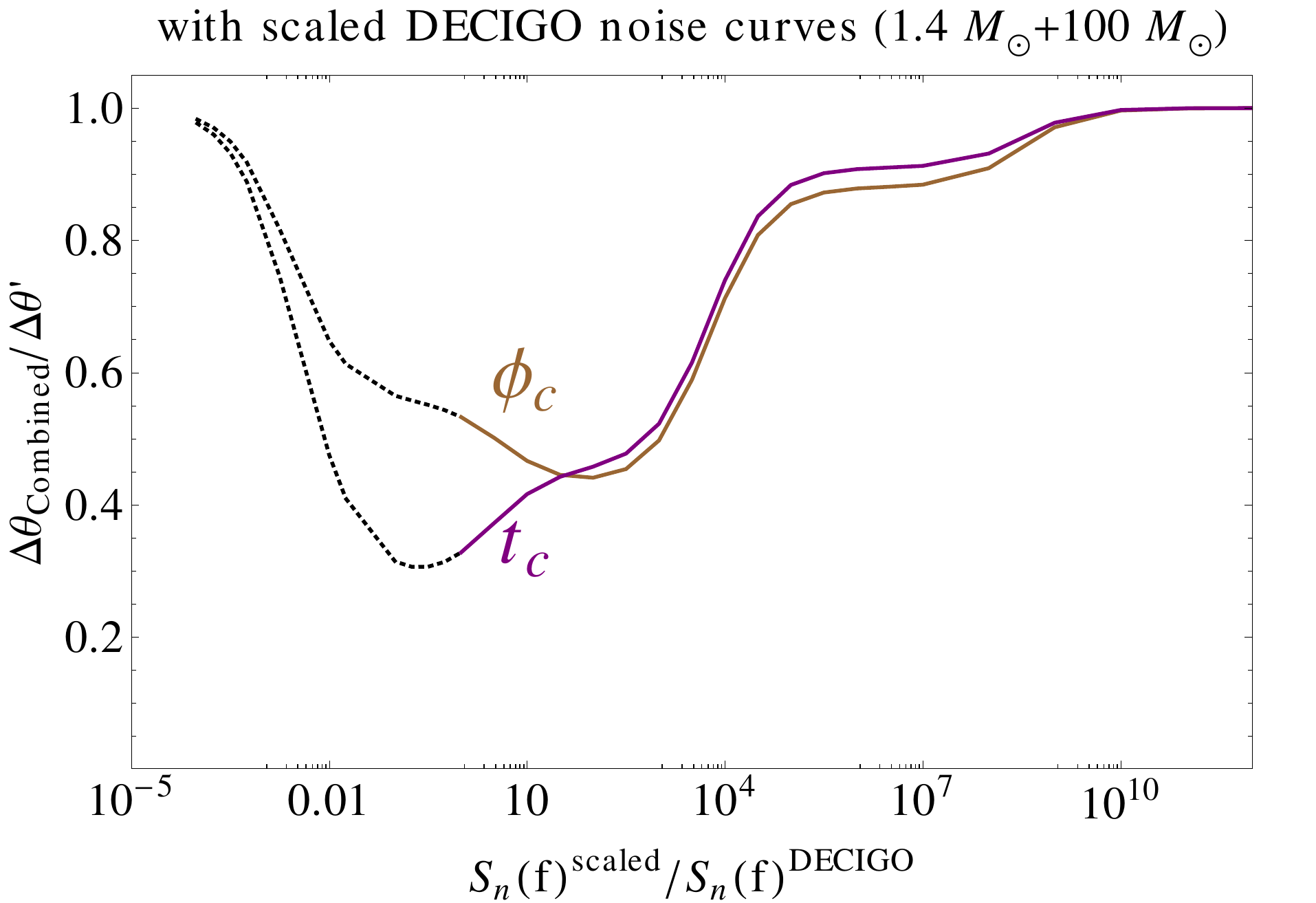}}
\caption{Variation in the error estimates on the chirp mass and the 
symmetric mass ratio (a), and the time and phase at coalescence (b), with 
varying DECIGO sensitivities. $\Delta \theta_{\rm comb}$
($\theta={\cal M}/\nu/t_c/p_c$) is obtained as explained in section \ref{secdata}, 
and $\Delta \theta '$ is as given in equation (\ref{thetap}). Here too, the dotted curves 
correspond to better sensitivity than DECIGO.}
\label{figmcnu3}
\end{figure}
We further analyse the variation in the error estimates on the parameters with the variation of 
the DECIGO sensitivity according to equation (\ref{sdecigo}). 
As a reference, we introduce a quantity $\Delta\theta '$ 
for a parameter $\theta$, which corresponds to the error estimate 
from the combined measurement if that single parameter 
was being estimated (equation \eqref{thetap}). 
In such a case, the joint probability for the 
combined estimate of some parameter say $\theta$ is given by
\begin{eqnarray*}
P(\theta) &\propto& e^{-(\Delta\theta)^2/(2 (\Delta\theta_1)^2)}~\times ~
e^{-(\Delta\theta)^2/(2 (\Delta\theta_2)^2)}\\
 &=& e^{-(\Delta\theta)^2/(2 (\Delta\theta ')^2)},
\end{eqnarray*}
where $\Delta\theta_1$ and $\Delta\theta_2$ are the error estimates from the 
first and the second measurements, respectively, and  
\begin{equation}
\Delta \theta '=\frac{\Delta \theta_ 1 \Delta \theta_ 2}
{\sqrt{\Delta \theta_ 1 ^2+\Delta \theta _2 ^2}}.
\label{thetap}
\end{equation}

These reference values are, off course, different from the correct estimate 
$\Delta\theta(\rm combined)=$ $\sqrt{C_{\rm comb}^{\theta\theta}}$ 
based on $C_{\rm comb}$ given in equation~\eqref{Ccomb}. 
The correct estimates of errors normalized by these reference values 
are shown in figure (\ref{figmcnu2}) and (\ref{figmcnu3}).
From our study on the synergy between (scaled-)DECIGO and ET, we can 
infer the following about the PN template parameters: for the chirp mass, there is a wide range of
sensitivities in the pre-DECIGO phase (sensitivities that are 
~$10$ - ~$10^7$ order of magnitude worse than DECIGO) where there is substantial gain
in combining the space-ground measurements, for the symmetric mass ratio
and the phase at coalescence maximum gain is obtained when the pre-DECIGO  sensitivity is almost $10^2$ order of magnitude worse than DECIGO,
and for the time at coalescence maximum gain is achieved in the
post-DECIGO phase where the sensitivity is almost an order of magnitude better than DECIGO.
\subsection{ppE expansion}
We present the results of the analysis for 1PN, 2PN and 3PN 
modifications to GR in this section. Here 1PN, 2PN and 3PN 
modifications mean that we fix $b=-1$, $b=-1/3$ and $b=1/3$ 
in equation (\ref{ppe}), respectively. 
The 1PN correction corresponds to the traditional Massive graviton 
theory (\cite{ppn1}), and 
the 2PN correction corresponds to Quadratic curvature 
theory (\cite{ppn1}). 
With distance fixed at 200 Mpc the results are shown in 
table \ref{tab3} and \ref{tab4} for 1PN, 
table \ref{tab5} and \ref{tab6} for 2PN 
and table \ref{tab7} and \ref{tab8} for 3PN. 
In all cases we find some gain to the constraint on 
these modifications to GR when combining
the measurements of the space and ground based detectors. 
\begin{figure}[h]
\centering \subfloat[Part 3][]{
\includegraphics[width=3.5in,height=2.5in]{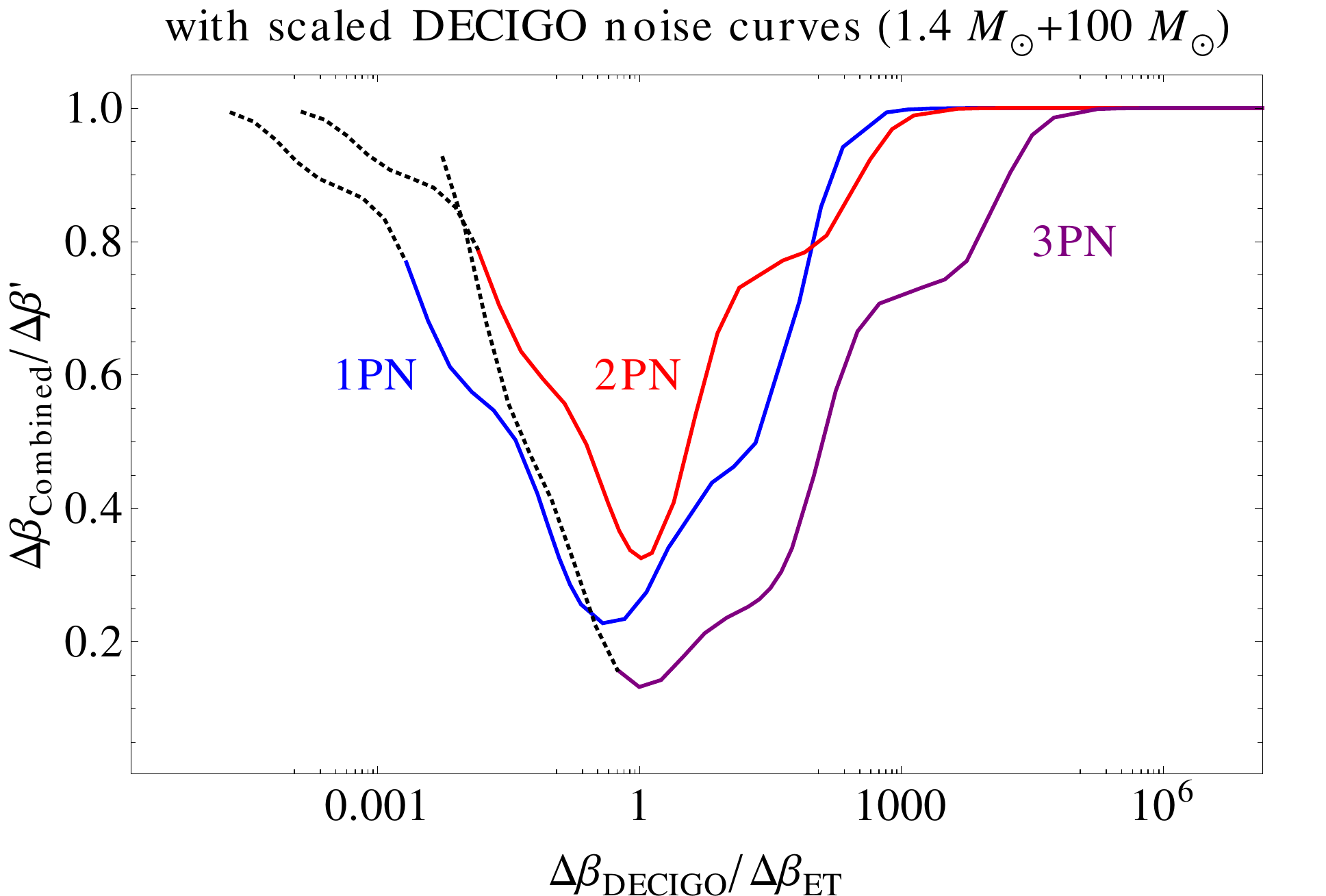}
\label{dhk1}
} ~~~~~~~
\subfloat[Part3][]{\includegraphics[width=3.6in,height=2.5in]{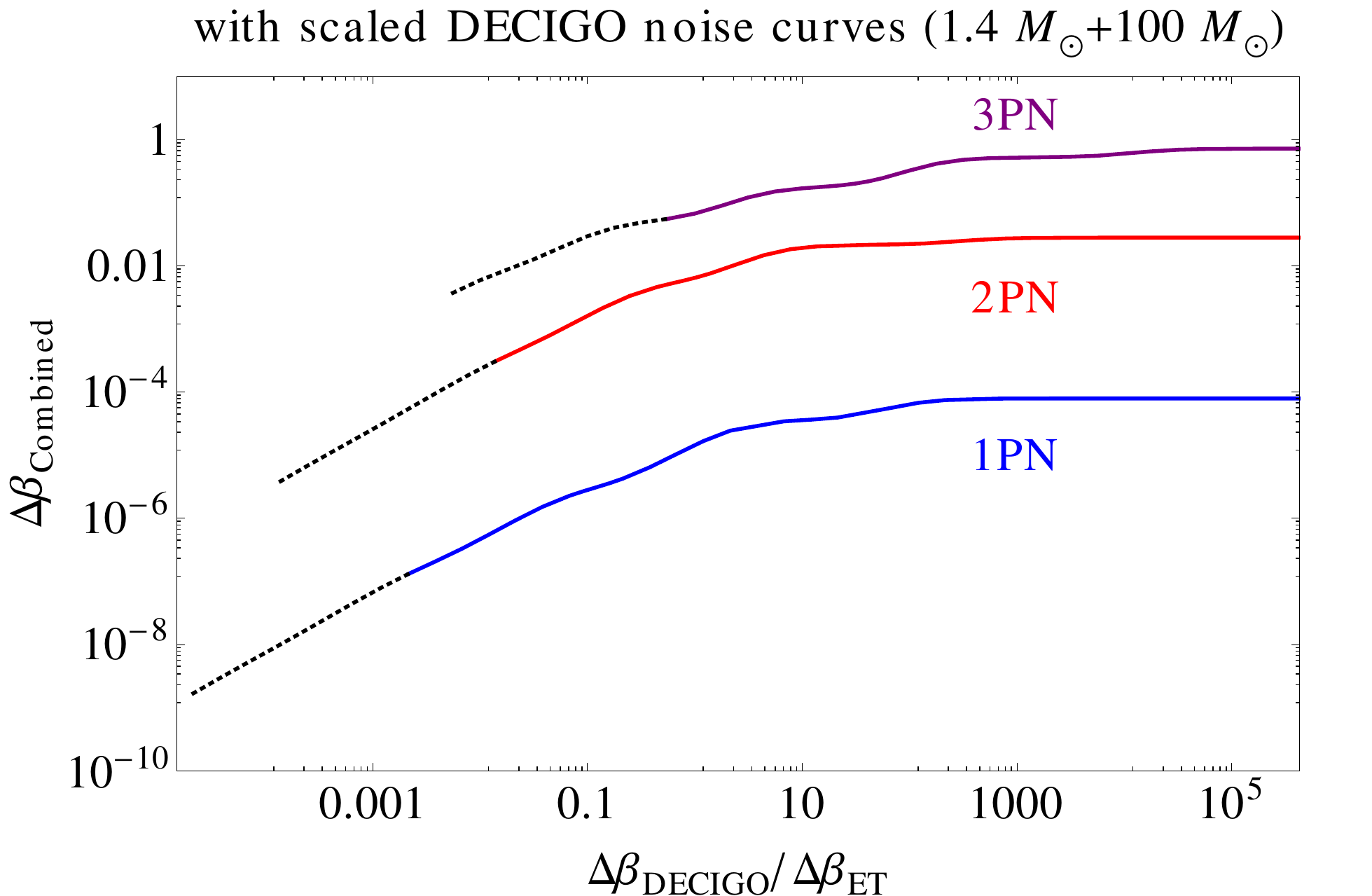}
\label{dhk2}}
\caption{Variation in the error estimates on the GR modification parameter at different 
PN orders. $\Delta \beta$ (combined) is obtained as explained in section \ref{secdata}, 
and $\Delta \beta '$ is as given in equation (\ref{thetap}). 
The dashed part corresponds to the noise curves with 
better sensitivity than DECIGO}
\label{dhk}
\end{figure}
\begin{figure}[h]
\centering \subfloat[Part 3][]{
\includegraphics[width=3.5in,height=2.5in]{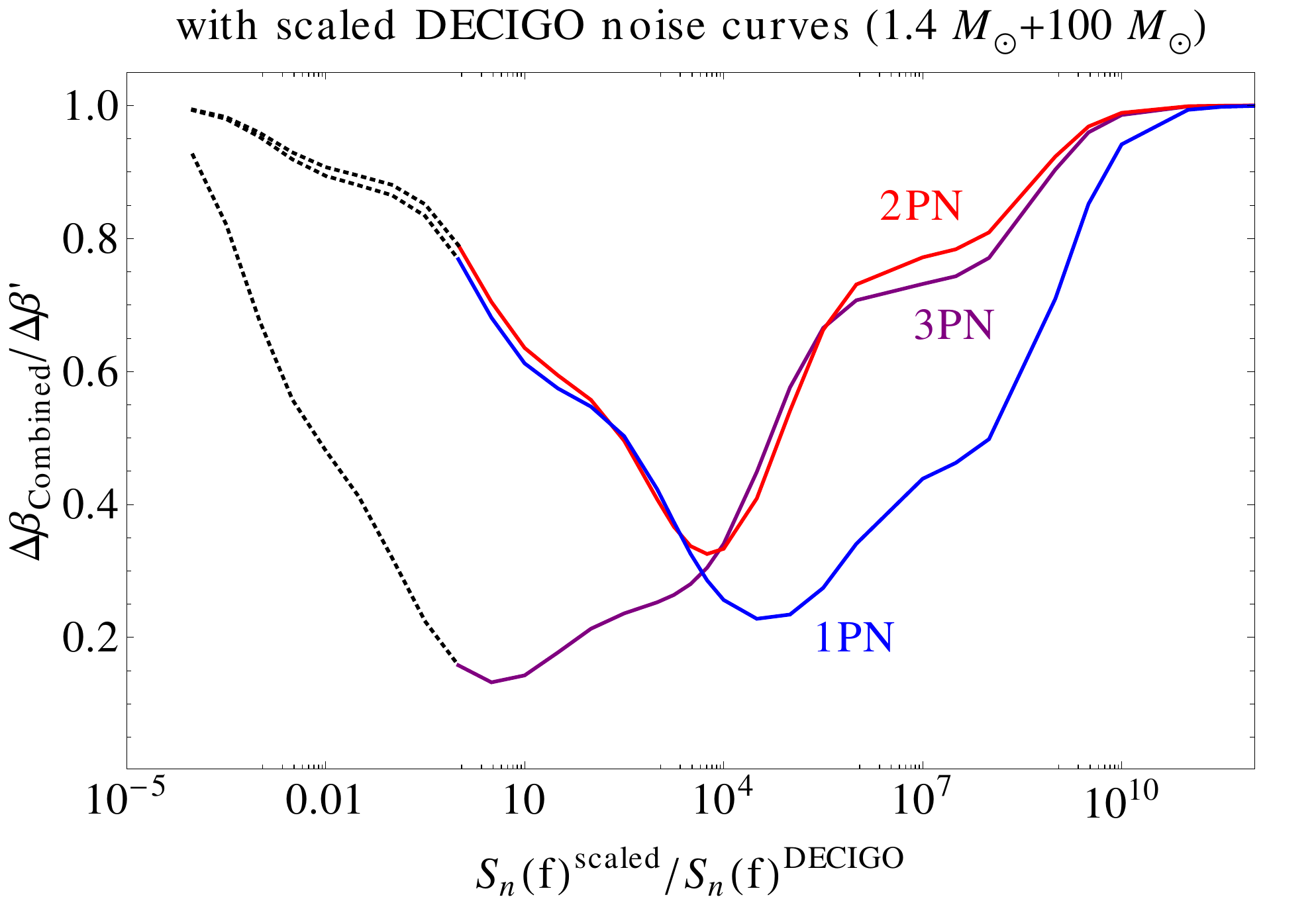}} 
~~~~~~~\subfloat[Part3][]{\includegraphics[width=3.6in,height=2.5in]
{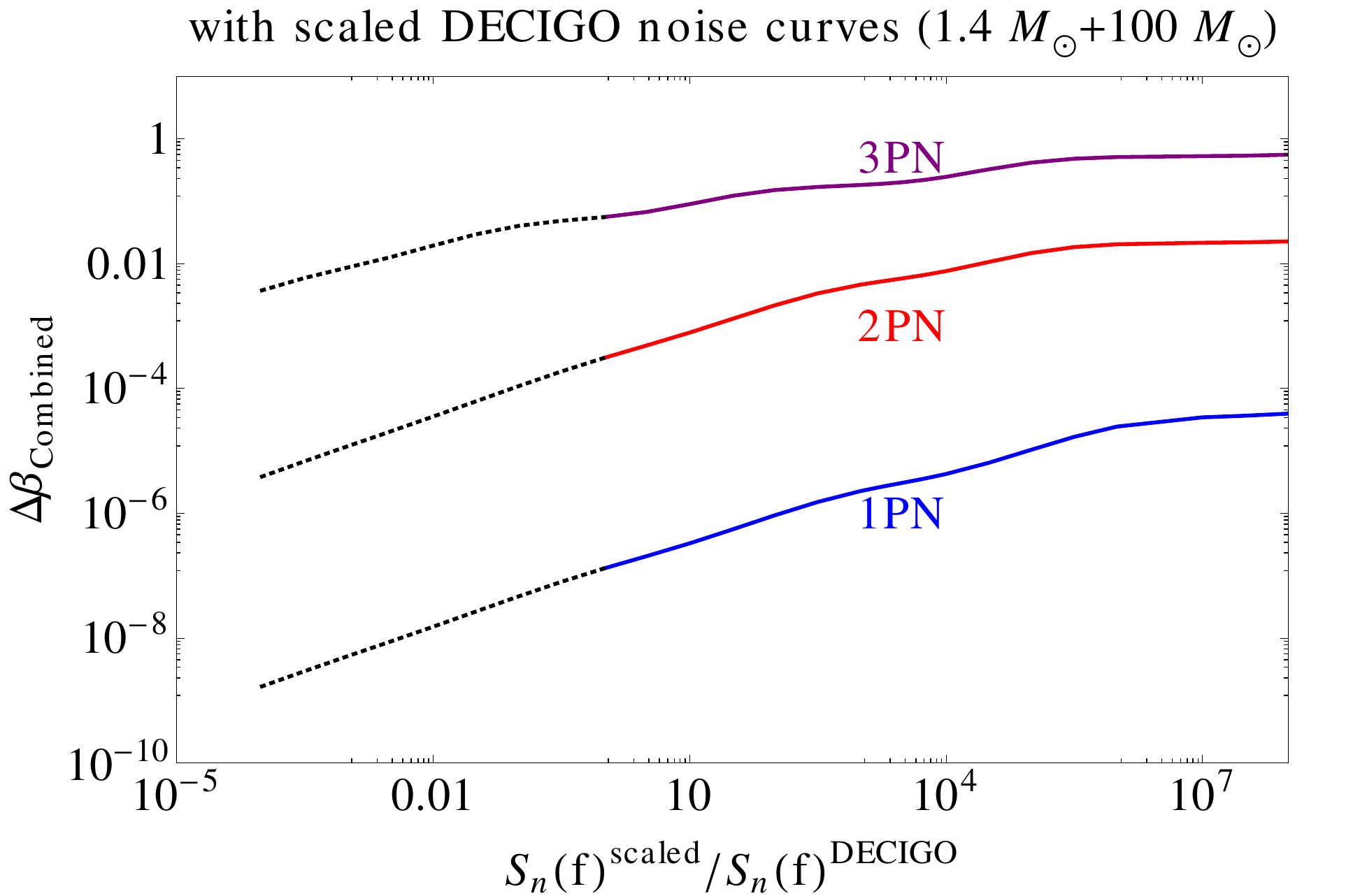}}
\caption{Variation in the error estimates on the GR modification parameter
with varying pre-DECIGO sensitivity. $\Delta \beta$ (combined)
is obtained as explained in section \ref{secdata}, 
and $\Delta \beta '$ is as given in equation (\ref{thetap}). 
The dotted curves correspond to better sensitivity 
than DECIGO.}
\label{dhl}
\end{figure}
In order to clarify the synergy effect, 
we present the plots similar to \ref{figmcnu2} and \ref{figmcnu3} 
for the parameters of the magnitude of modification from GR. 
As before, in figure (\ref{dhk})
we plot the error estimate based on $C_{\rm comb}$ given in equation~\eqref{Ccomb}, $\Delta\beta$, 
relative to the reference value $\Delta\beta'$ defined by 
equation \eqref{thetap}, for combined measurements from scaled-DECIGO and ET.
For very low DECIGO sensitivity the error estimates are dominated by 
ET, while for very high DECIGO sensitivity they are dominated by
DECIGO. In both limiting cases we do not have any merit of combining 
the space-ground measurements. 
However, in the middle range we have a sweet spot where the gain becomes
significantly large. In this plot (figure (\ref{dhk})) the horizontal axis is
$\Delta\beta_{\rm DECIGO}/\Delta\beta_{\rm ET}$. 
For all PN order modifications, the sweet spot appears around 
$\Delta\beta_{\rm DECIGO}/\Delta\beta_{\rm ET}=1$, 
where both 
detectors equally contribute to constrain the parameter $\beta$. 
In figure (\ref{dhk2}) we plot $\Delta\beta _{Combined}$. 
In the region with
$\Delta\beta_{\rm DECIGO}/\Delta\beta_{\rm ET}\ll 1$, 
the error estimates are dominated by the DECIGO 
sensitivity, and hence they show linear dependence. 
In the region with
$\Delta\beta_{\rm DECIGO}/\Delta\beta_{\rm ET}\gg 1$, 
the error estimates are dominated by the ET 
sensitivity, and hence the curves become almost flat. 
At around $\Delta\beta_{\rm DECIGO}/\Delta\beta_{\rm ET}= 1$, 
the slope of the curves changes. This change is 
not monotonic, and corresponds to the sweet spot in the figure 
(\ref{dhk1}) where the true joint estimates 
are better than the reference estimates given by $\Delta \beta'$. 

We also show these error estimates as functions of 
the sensitivity level of the scaled-DECIGO in figure (\ref{dhl}). 
The higher order post-Newtonian terms tend to be better determined 
by the ground based detectors which have better sensitivities 
at higher frequencies. Hence, for higher PN order terms, the synergy
is maximum (i.e. the space and ground detectors contribute almost
equally) for higher sensitivity of scaled-DECIGO.
The positions of the sweet spots in figure (\ref{figmcnu3}) can be understood 
in the same way by noticing that the leading order frequency dependences
of the chirp mass, the symmetric mass ratio, and the phase and time at
coalescence are 0PN, 1PN, 2.5PN and 4PN, respectively. 
\section{Conclusions}\label{conclusion}
In this paper we have assessed the expected synergy effects 
between ground and space based detectors on the parameter estimation
of coalescing binary systems. 
Our aim is to demonstrate that the advantage of having a GW antenna 
which is sensitive at low frequencies, is larger than we naively expect.  
Larger gain in the error estimate of the GW waveform parameters is obtained when the constraints from individual ground or space detectors are almost equal. In case of the ppE parameters that characterize deviations from 
GR, some gain is always obtained irrespective of the level of sensitivity of the GW antenna. For lower PN order modifications some gain can be obtained even if the sensitivity of the space detector is degraded. 
The same argument also applies to the estimated errors in 
the extraction of binary parameters contained in the standard 
PN templates. In the present paper, neglecting the spins, we 
considered the chirp mass, the symmetric mass ratio, and 
the phase and time at coalescence, whose leading order 
frequency dependences on the GW phase are 0PN, 1PN, 2.5PN and 
4PN, respectively. We found a sweet spot in the combined 
estimates of the chirp mass and the symmetric mass ratio 
in the pre-DECIGO phase (worse sensitivity than DECIGO).
In this region, the combined error estimates are better 
than individual detector estimates. Substantial gain was 
also obtained in the joint measurement of the time and phase 
at coalescence. Here too the sweet spot was in the 
pre-DECIGO phase for the phase at coalescence, but for the time at coalescence, maximum gain was obtained with post-DECIGO sensitivity. 
As expected, for higher PN order parameters, synergy is obtained 
at higher sensitivities of the space-borne GW detector.

The sensitivity of space-borne GW antenna such as pre-DECIGO, 
which is the precursor mission for DECIGO, is still to be determined 
by considering various aspects, including the scientific gain. 
The study of the synergy effects reported in the present paper provides valuable information and can be taken into account to make such decisions.

\begin{acknowledgments}
This work was supported by MEXT Grant-in-Aid for
Scientific Research on Innovative Areas, ``New Developments
in Astrophysics Through Multi-Messenger Observations
of Gravitational Wave Sources'', Nos. 24103001 and 
24103006. This work was also supported in part by
Grant-in-Aid for Scientific Research (B) No. 26287044. 
R.N acknowledges support under CSIR-SRF scheme (Govt. of India), and
thanks the Department of Physics, Kyoto University where part of this work was done, for hospitality. S.J acknowledges support from grant under ISRO-RESPOND program (ISRO/RES/2/384/2014-15).
\end{acknowledgments}

\appendix*
\section{Tabels}
\begin{table}[h]
\caption{Errors estimates for PN waveform parameters for individual detector measurements.
These are calculated for NS-BH binaries with masses $1.4~M_{\odot}$+$10~M_{\odot}$ and
$1.4~M_{\odot}$+$100~M_{\odot}$. The distance is fixed to 200 Mpc.}
\vspace{0.2cm}
\centering
\begin{tabular}{|c c c c c c|}
\hline {\footnotesize  Binary mass} & {\footnotesize  Detector} & 
{\footnotesize $\Delta t_c$} & {\footnotesize $\Delta \phi_c$ } &  
{\footnotesize  $\Delta {\cal M} /{\cal M} (\%)$ } &  {\footnotesize  $\Delta \nu / \nu (\%)$} 
\\
\hline

$1.4 M_{\odot}$+$10 M_{\odot}$ \quad & DECIGO  \quad & $9.16 \times 10^{-5}$ 
\quad& $9.70 \times 10^{-4}$ \quad& $9.57 \times 10^{-8}$ 
\quad& $2.60 \times 10^{-4}$ \quad\\
$1.4 M_{\odot}$+$10 M_{\odot}$ \quad & eLISA  \quad & $1.30 \times 10^2$ \quad& $3.87 \times 10^2$ 
\quad& $8.16 \times 10^{-3}$ \quad& $2.80 \times 10^{1}$ \quad\\
$1.4 M_{\odot}$+$10 M_{\odot}$ \quad & advLIGO  \quad & $3.22 \times 10^{-4}$ 
\quad& $3.26 \times 10^{-1}$ \quad& $1.62 \times 10^{-2}$ \quad& $4.52
\times 10^{-1}$ \quad\\
$1.4 M_{\odot}$+$10 M_{\odot}$ \quad & ET  \quad & $1.51 \times 10^{-5}$ 
\quad& $1.64 \times 10^{-2}$ \quad& $1.35 \times 10^{-4}$ \quad& $1.31 \times 10^{-2}$ 
\quad\\
\hline
\hline
$1.4 M_{\odot}$+$100 M_{\odot}$ \quad & DECIGO  \quad & $6.52 \times 10^{-5}$ 
\quad& $2.43 \times 10^{-3}$ \quad& $7.36 \times 10^{-8}$ 
\quad& $9.36 \times 10^{-5}$ \quad\\
$1.4 M_{\odot}$+$100 M_{\odot}$ \quad & eLISA  \quad & $1.08 \times 10^2$ \quad& $2.07 \times 10^3$ 
\quad& $6.20 \times 10^{-2}$ \quad& $6.73 \times 10^{1}$ \quad\\
$1.4 M_{\odot}$+$100 M_{\odot}$ \quad & advLIGO  \quad & $2.09 \times 10^{-4}$ \quad& $1.49 \times 10^{-1}$ \quad& $3.43 \times 10^{-2}$ \quad& $3.53 \times 10^{-2}$ \quad\\
$1.4 M_{\odot}$+$100 M_{\odot}$ \quad & ET  \quad & $1.02 \times 10^{-5}$ 
\quad& $6.02 \times 10^{-3}$ \quad& $3.00 \times 10^{-4}$ 
\quad& $1.23 \times 10^{-3}$ \quad\\
\hline
\end{tabular}
\label{tab1}
\end{table}
\begin{table}[h]
  \caption{Errors estimates for PN waveform parameters for combined detector measurements.
These are calculated for NS-BH binaries with masses $1.4~M_{\odot}$+$10~M_{\odot}$ and
$1.4~M_{\odot}$+$100~M_{\odot}$. The distance is fixed to 200 Mpc.}
  \vspace{0.2cm}
  \centering
  \begin{tabular}{|c c c c c c|}
    \hline {\footnotesize
  Binary mass} & {\footnotesize
  Detector} & {\footnotesize $\Delta t_c$} & {\footnotesize $\Delta \phi_c$ } &  
  {\footnotesize  $\Delta {\cal M} /{\cal M} (\%)$ } &  {\footnotesize  $\Delta \nu / \nu (\%)$ } \\
    \hline
    
$1.4 M_{\odot}$+$10 M_{\odot}$ \quad & ET+DECIGO \quad & $6.04 \times 10^{-6}$ 
\quad& $8.98 \times 10^{-4}$ \quad& $9.15 \times 10^{-8}$ 
\quad& $2.46 \times 10^{-4}$ \quad\\
$1.4 M_{\odot}$+$10 M_{\odot}$ \quad & ET+eLISA \quad & $1.47 \times 10^{-5}$ 
\quad& $1.58 \times 10^{-2}$ \quad& $1.03 \times 10^{-4}$ 
\quad& $1.21 \times 10^{-2}$ \quad\\
$1.4 M_{\odot}$+$10 M_{\odot}$ \quad & advLIGO+DECIGO \quad & $4.95 \times 10^{-5}$ 
\quad& $9.22 \times 10^{-4}$ \quad& $9.28 \times 10^{-8}$ 
\quad& $2.50 \times 10^{-4}$ \quad\\
$1.4 M_{\odot}$+$10 M_{\odot}$ \quad & advLIGO+eLISA \quad & $2.36 \times 10^{-4}$ 
\quad& $2.55 \times 10^{-1}$ \quad& $1.79 \times 10^{-4}$ \quad& $2.41 \times 10^{-1}$ \quad\\
 \hline
  \hline
$1.4 M_{\odot}$+$100 M_{\odot}$ \quad & ET+DECIGO \quad & $3.31 \times 10^{-6}$ 
\quad& $1.20 \times 10^{-3}$ \quad& $3.92 \times 10^{-8}$ 
\quad& $5.24 \times 10^{-5}$ \quad\\
$1.4 M_{\odot}$+$100 M_{\odot}$ \quad & ET+eLISA \quad & $9.84 \times 10^{-6}$ 
\quad& $5.73 \times 10^{-3}$ \quad& $2.28 \times 10^{-4}$ 
\quad& $1.14 \times 10^{-3}$ \quad\\
$1.4 M_{\odot}$+$100 M_{\odot}$ \quad & advLIGO+DECIGO \quad & $2.56 \times 10^{-5}$ 
\quad& $1.56 \times 10^{-3}$ \quad& $4.90 \times 10^{-8}$ 
\quad& $6.52 \times 10^{-5}$ \quad\\
$1.4 M_{\odot}$+$100 M_{\odot}$ \quad & advLIGO+eLISA \quad & $1.74 \times 10^{-4}$ 
\quad& $1.36 \times 10^{-1}$ \quad& $3.53 \times 10^{-4}$ 
\quad& $2.22 \times 10^{-2}$ \quad\\
 \hline  
  \end{tabular}
  \label{tab2}
\end{table}
\begin{table}[h]
  \caption{Errors estimates for ppE waveform parameters for individual detector 
  measurements, when the modification to GR appears at the first PN order. 
  These are calculated for NS-BH binaries with masses 
 $1.4~M_{\odot}$+$100~M_{\odot}$. The distance is fixed 
  to 200 Mpc.}
  \vspace{0.2cm}
  \centering
  \begin{tabular}{|c c c c c c c|}
    \hline {\footnotesize
  Binary mass} & {\footnotesize
  Detector} & {\footnotesize $\Delta t_c$} & {\footnotesize $\Delta \phi_c$ } &  
  {\footnotesize  $\Delta {\cal M} /{\cal M} (\%)$ } &  {\footnotesize  $\Delta \nu / \nu (\%)$} & 
  {\footnotesize $\Delta \beta$} \\
    \hline

$1.4 M_{\odot}$+$10 M_{\odot}$ \quad & DECIGO  \quad & $1.56 \times 10^{-4}$ 
\quad& $9.37 \times 10^{-3}$ \quad& $1.85 \times 10^{-7}$ \quad& $9.78 \times 10^{-4}$ 
\quad& $2.48 \times 10^{-7}$\quad\\
$1.4 M_{\odot}$+$10 M_{\odot}$ \quad & eLISA \quad & $2.59 \times 10^2$ \quad& $4.91 \times 10^3$ 
\quad& $2.98 \times 10^{-2}$ \quad& $3.00 \times 10^2$ \quad& $1.15 \times 10^{-1}$\quad\\
$1.4 M_{\odot}$+$10 M_{\odot}$ \quad & aLIGO  \quad & $7.16 \times 10^{-4}$ 
\quad& $7.09 \times 10^{-1}$ \quad& $9.13 \times 10^{-2}$ \quad& $2.33$ \quad& $8.48 \times 10^{-3}$\quad\\
$1.4 M_{\odot}$+$10 M_{\odot}$ \quad & ET  \quad & $2.33 \times 10^{-5}$ \quad& $1.75 \times 10^{-2}$ 
\quad& $4.64 \times 10^{-4}$ \quad& $4.44 \times 10^{-2}$ \quad& $8.86 \times 10^{-5}$\quad\\
\hline
\hline
$1.4 M_{\odot}$+$100 M_{\odot}$ \quad & DECIGO  \quad & $7.56 \times 10^{-5}$ 
\quad& $3.55 \times 10^{-3}$ \quad& $3.76 \times 10^{-7}$ \quad& $2.03 \times 10^{-4}$ 
\quad& $1.69 \times 10^{-7}$\quad\\
$1.4 M_{\odot}$+$100 M_{\odot}$ \quad & eLISA \quad & $1.22 \times 10^2$ \quad& $2.12 \times 10^3$ 
\quad& $9.18 \times 10^{-3}$ \quad& $6.74 \times 10^{1}$ 
\quad& $1.26 \times 10^{-2}$\quad\\
$1.4 M_{\odot}$+$100 M_{\odot}$ \quad & aLIGO \quad & $7.18 \times 10^{-4}$ 
\quad& $1.93$ \quad& $1.32 \times 10^{-1}$ \quad& $1.16 \times 10^{-1}$ 
\quad& $5.46 \times 10^{-2}$\quad\\
$1.4 M_{\odot}$+$100 M_{\odot}$ \quad & ET \quad & $2.01 \times 10^{-5}$ 
\quad& $3.42 \times 10^{-2}$ \quad& $7.98 \times 10^{-4}$ \quad& $3.36 \times 10^{-3}$ 
\quad& $7.89 \times 10^{-5}$\quad\\
\hline
 \hline
  \end{tabular}
    \label{tab3}
\end{table}
\begin{table}[h]
  \caption{Errors estimates for ppE waveform parameters for combined detector 
  measurements, when the modification to GR appears at the first PN order. 
  These are calculated for NS-BH binaries with masses 
 $1.4~M_{\odot}$+$100~M_{\odot}$. The distance is fixed 
  to 200 Mpc.}
  \vspace{0.2cm}
  \centering
  \begin{tabular}{|c c c c c c c|}
    \hline {\footnotesize
  Binary mass} & {\footnotesize
  Detector} & {\footnotesize $\Delta t_c$} & {\footnotesize $\Delta \phi_c$ } &  {\footnotesize  $\Delta {\cal M} /{\cal M} (\%)$ } &  {\footnotesize  $\Delta \nu / \nu (\%)$} & {\footnotesize $\Delta \beta$} \\
    \hline

$1.4 M_{\odot}$+$10 M_{\odot}$ \quad & ET+DECIGO  \quad & $7.43 \times 10^{-6}$ \quad& $4.54 \times 10^{-3}$ \quad& $1.33 \times 10^{-7}$ \quad& $5.84 \times 10^{-4}$ \quad& $1.26 \times 10^{-7}$\quad\\
$1.4 M_{\odot}$+$10 M_{\odot}$ \quad & ET+eLISA \quad & $1.91 \times 10^{-5}$ \quad& $1.72 \times 10^{-2}$ \quad& $1.82 \times 10^{-4}$ \quad& $2.86 \times 10^{-2}$ \quad& $4.50 \times 10^{-5}$\quad\\
$1.4 M_{\odot}$+$10 M_{\odot}$ \quad & aLIGO+DECIGO  \quad & $5.71 \times 10^{-5}$ \quad& $6.12 \times 10^{-3}$ \quad& $1.55 \times 10^{-7}$ \quad& $7.32 \times 10^{-4}$ \quad& $1.67 \times 10^{-7}$\quad\\
$1.4 M_{\odot}$+$10 M_{\odot}$ \quad & aLIGO+eLISA \quad & $3.3 \times 10^{-4}$ \quad& $2.60 \times 10^{-1}$ \quad& $1.10 \times 10^{-3}$ \quad& $6.10 \times 10^{-1}$ \quad& $1.13 \times 10^{-3}$\quad\\
\hline
\hline
$1.4 M_{\odot}$+$100 M_{\odot}$ \quad & ET+DECIGO  \quad & $3.87 \times 10^{-6}$ \quad& $1.47 \times 10^{-3}$ \quad& $2.53 \times 10^{-7}$ \quad& $1.09 \times 10^{-4}$ \quad& $1.30 \times 10^{-7}$\quad\\
$1.4 M_{\odot}$+$100 M_{\odot}$ \quad & ET+eLISA \quad & $1.59 \times 10^{-5}$ \quad& $2.23 \times 10^{-2}$ \quad& $3.66 \times 10^{-4}$ \quad& $2.50 \times 10^{-3}$ \quad& $4.72 \times 10^{-5}$\quad\\
$1.4 M_{\odot}$+$100 M_{\odot}$ \quad & aLIGO+DECIGO  \quad & $2.67 \times 10^{-5}$ \quad& $2.15 \times 10^{-3}$ \quad& $3.09 \times 10^{-7}$ \quad& $1.46 \times 10^{-4}$ \quad& $1.51 \times 10^{-7}$\quad\\
$1.4 M_{\odot}$+$100 M_{\odot}$ \quad & aLIGO+eLISA \quad & $3.01 \times 10^{-4}$ \quad& $4.95 \times 10^{-1}$ \quad& $1.85 \times 10^{-3}$ \quad& $5.14 \times 10^{-2}$ \quad& $1.21 \times 10^{-3}$\quad\\
\hline
 \hline
  \end{tabular}
    \label{tab4}
\end{table}
\begin{table}[h]
  \caption{Errors estimates for ppE waveform parameters for individual detector 
  measurements, when the modification to GR appears at the second PN order. 
  These are calculated for NS-BH binaries with masses 
  $1.4~M_{\odot}$+$100~M_{\odot}$. The distance is fixed 
  to 200 Mpc.}
  \vspace{0.2cm}
  \centering
  \begin{tabular}{|c c c c c c c|}
    \hline {\footnotesize
  Binary mass} & {\footnotesize
  Detector} & {\footnotesize $\Delta t_c$} & {\footnotesize $\Delta \phi_c$ } &  {\footnotesize  $\Delta {\cal M} /{\cal M} (\%)$ } &  {\footnotesize  $\Delta \nu / \nu (\%)$} & {\footnotesize $\Delta \beta$} \\
    \hline

$1.4 M_{\odot}$+$10 M_{\odot}$ \quad & DECIGO  \quad & $1.63 \times 10^{-4}$ \quad& $1.27 \times 10^{-2}$ \quad& $1.66 \times 10^{-7}$ \quad& $5.62 \times 10^{-4}$ \quad& $3.51 \times 10^{-4}$\quad\\
$1.4 M_{\odot}$+$10 M_{\odot}$ \quad & eLISA \quad & $2.76 \times 10^2$ \quad& $7.27 \times 10^3$ \quad& $2.47 \times 10^{-2}$ \quad& $1.17 \times 10^{2}$ \quad& $1.84 \times 10^2$\quad\\
$1.4 M_{\odot}$+$10 M_{\odot}$ \quad & aLIGO  \quad & $3.95 \times 10^{-3}$ \quad& $8.21 \times 10^1$ \quad& $7.15 \times 10^{-2}$ \quad& $4.21 \times 10^1$ \quad& $1.60 \times 10^1$\quad\\
$1.4 M_{\odot}$+$10 M_{\odot}$ \quad & ET  \quad & $1.00 \times 10^{-4}$ \quad& $1.90$ \quad& $1.21 \times 10^{-3}$ \quad& $9.95 \times 10^{-1}$ \quad& $3.72 \times 10^{-1}$\quad\\
\hline
\hline
$1.4 M_{\odot}$+$100 M_{\odot}$ \quad & DECIGO  \quad & $7.45 \times 10^{-5}$ \quad& $2.53 \times 10^{-3}$ \quad& $4.40 \times 10^{-7}$ \quad& $3.57 \times 10^{-4}$ \quad& $3.88 \times 10^{-4}$\quad\\
$1.4 M_{\odot}$+$100 M_{\odot}$ \quad & eLISA \quad & $1.22 \times 10^2$ \quad& $2.20 \times 10^3$ \quad& $6.26 \times 10^{-2}$ \quad& $6.76 \times 10^{1}$ \quad& $2.75 \times 10^1$\quad\\
$1.4 M_{\odot}$+$100 M_{\odot}$ \quad & aLIGO \quad & $1.18 \times 10^{-3}$ \quad& $7.02$ \quad& $6.25 \times 10^{-2}$ \quad& $3.19 \times 10^{-1}$ \quad& $9.05 \times 10^{-1}$\quad\\
$1.4 M_{\odot}$+$100 M_{\odot}$ \quad & ET \quad & $4.42 \times 10^{-5}$ \quad& $2.22 \times 10^{-1}$ \quad& $4.08 \times 10^{-4}$ \quad& $1.14 \times 10^{-2}$ \quad& $2.78 \times 10^{-2}$\quad\\

\hline
 \hline
  \end{tabular}
    \label{tab5}
\end{table}
\begin{table}[h]
  \caption{Errors estimates for ppE waveform parameters for combined detector 
  measurements, when the modification to GR appears at the second PN order. 
  These are calculated for NS-BH binaries with masses 
  $1.4~M_{\odot}$+$100~M_{\odot}$. The distance is fixed 
  to 200 Mpc.}
  \vspace{0.2cm}
  \centering
  \begin{tabular}{|c c c c c c c|}
    \hline {\footnotesize
  Binary mass} & {\footnotesize
  Detector} & {\footnotesize $\Delta t_c$} & {\footnotesize $\Delta \phi_c$ } &  {\footnotesize  $\Delta {\cal M} /{\cal M} (\%)$ } &  {\footnotesize  $\Delta \nu / \nu (\%)$} & {\footnotesize $\Delta \beta$} \\
    \hline

$1.4 M_{\odot}$+$10 M_{\odot}$ \quad & ET+DECIGO  \quad & $7.64 \times 10^{-6}$ \quad& $5.80 \times 10^{-3}$ \quad& $1.21 \times 10^{-7}$ \quad& $3.76 \times 10^{-4}$ \quad& $1.66 \times 10^{-4}$\quad\\
$1.4 M_{\odot}$+$10 M_{\odot}$ \quad & ET+eLISA \quad & $3.27 \times 10^{-5}$ \quad& $4.88 \times 10^{-1}$ \quad& $2.50 \times 10^{-4}$ \quad& $2.62 \times 10^{-1}$ \quad& $9.65 \times 10^{-2}$\quad\\
$1.4 M_{\odot}$+$10 M_{\odot}$ \quad & aLIGO+DECIGO  \quad & $5.77 \times 10^{-5}$ \quad& $8.09 \times 10^{-3}$ \quad& $1.41 \times 10^{-7}$ \quad& $4.54 \times 10^{-4}$ \quad& $2.28 \times 10^{-4}$\quad\\
$1.4 M_{\odot}$+$10 M_{\odot}$ \quad & aLIGO+eLISA \quad & $5.20 \times 10^{-4}$ \quad& $8.23$ \quad& $1.83 \times 10^{-3}$ \quad& $4.37$ \quad& $1.61$\quad\\
\hline
\hline
$1.4 M_{\odot}$+$100 M_{\odot}$ \quad & ET+DECIGO  \quad & $4.15 \times 10^{-6}$ \quad& $2.04 \times 10^{-3}$ \quad& $3.13 \times 10^{-7}$ \quad& $2.36 \times 10^{-4}$ \quad& $3.06 \times 10^{-4}$\quad\\
$1.4 M_{\odot}$+$100 M_{\odot}$ \quad & ET+eLISA \quad & $4.03 \times 10^{-5}$ \quad& $1.96 \times 10^{-1}$ \quad& $2.73 \times 10^{-4}$ \quad& $1.03 \times 10^{-2}$ \quad& $2.44 \times 10^{-2}$\quad\\
$1.4 M_{\odot}$+$100 M_{\odot}$ \quad & aLIGO+DECIGO  \quad & $2.67 \times 10^{-5}$ \quad& $2.12 \times 10^{-3}$ \quad& $3.74 \times 10^{-7}$ \quad& $2.92 \times 10^{-4}$ \quad& $3.52 \times 10^{-4}$\quad\\
$1.4 M_{\odot}$+$100 M_{\odot}$ \quad & aLIGO+eLISA \quad & $7.57 \times 10^{-4}$ \quad& $3.91$ \quad& $5.42 \times 10^{-4}$ \quad& $1.98 \times 10^{-1}$ \quad& $4.97 \times 10^{-1}$\quad\\
\hline
 \hline
  \end{tabular}
    \label{tab6}
\end{table}
\begin{table}[h]
  \caption{Errors estimates for ppE waveform parameters for individual detector 
  measurements, when the modification to GR appears at the third PN order. 
  These are calculated for NS-BH binaries with masses 
  $1.4~M_{\odot}$+$100~M_{\odot}$. The distance is fixed 
  to 200 Mpc.}
  \vspace{0.2cm}
  \centering
  \begin{tabular}{|c c c c c c c|}
    \hline {\footnotesize
  Binary mass} & {\footnotesize
  Detector} & {\footnotesize $\Delta t_c$} & {\footnotesize $\Delta \phi_c$ } &  {\footnotesize  $\Delta {\cal M} /{\cal M} (\%)$ } &  {\footnotesize  $\Delta \nu / \nu (\%)$} & {\footnotesize $\Delta \beta$} \\
    \hline

$1.4 M_{\odot}$+$10 M_{\odot}$ \quad & DECIGO  \quad & $2.29 \times 10^{-4}$ \quad& $6.43 \times 10^{-3}$ \quad& $1.29 \times 10^{-7}$ \quad& $3.86 \times 10^{-4}$ \quad& $2.48 \times 10^{-1}$\quad\\
$1.4 M_{\odot}$+$10 M_{\odot}$ \quad & eLISA \quad & $3.98 \times 10^2$ \quad& $4.68 \times 10^3$ \quad& $1.65 \times 10^{-2}$ \quad& $6.59 \times 10^{1}$ \quad& $2.08 \times 10^5$\quad\\
$1.4 M_{\odot}$+$10 M_{\odot}$ \quad & aLIGO  \quad & $1.19 \times 10^{-3}$ \quad& $7.69$ \quad& $2.93 \times 10^{-2}$ \quad& $2.44$ \quad& $4.67 \times 10^1$\quad\\
$1.4 M_{\odot}$+$10 M_{\odot}$ \quad & ET  \quad & $4.80 \times 10^{-5}$ \quad& $2.15 \times 10^{-1}$ \quad& $1.52 \times 10^{-4}$ \quad& $5.08 \times 10^{-2}$ \quad& $1.38$\quad\\
\hline
\hline
$1.4 M_{\odot}$+$100 M_{\odot}$ \quad & DECIGO  \quad & $1.8 \times 10^{-4}$ \quad& $3.24 \times 10^{-2}$ \quad& $6.83 \times 10^{-7}$ \quad& $6.75 \times 10^{-4}$ \quad& $4.02 \times 10^{-1}$\quad\\
$1.4 M_{\odot}$+$100 M_{\odot}$ \quad & eLISA \quad & $1.38 \times 10^2$ \quad& $2.50 \times 10^3$ \quad& $6.51 \times 10^{-2}$ \quad& $7.02 \times 10^{1}$ \quad& $3.07 \times 10^4$\quad\\
$1.4 M_{\odot}$+$100 M_{\odot}$ \quad & aLIGO \quad & $2.09 \times 10^{-4}$ \quad& $2.96$ \quad& $7.44 \times 10^{-2}$ \quad& $3.79 \times 10^{-1}$ \quad& $2.97 \times 10^1$\quad\\
$1.4 M_{\odot}$+$100 M_{\odot}$ \quad & ET \quad & $1.28 \times 10^{-5}$ \quad& $7.81 \times 10^{-2}$ \quad& $4.12 \times 10^{-4}$ \quad& $7.53 \times 10^{-3}$ \quad& $7.17 \times 10^{-1}$\quad\\

\hline
 \hline
  \end{tabular}
    \label{tab7}
\end{table}
\begin{table}[h]
  \caption{Errors estimates for ppE waveform parameters for combined detector 
  measurements, when the modification to GR appears at the third PN order. 
  These are calculated for NS-BH binaries with masses 
  $1.4~M_{\odot}$+$100~M_{\odot}$. The distance is fixed 
  to 200 Mpc.}
  \vspace{0.2cm}
  \centering
  \begin{tabular}{|c c c c c c c|}
    \hline {\footnotesize
  Binary mass} & {\footnotesize
  Detector} & {\footnotesize $\Delta t_c$} & {\footnotesize $\Delta \phi_c$ } &  {\footnotesize  $\Delta {\cal M} /{\cal M} (\%)$ } &  {\footnotesize  $\Delta \nu / \nu (\%)$} & {\footnotesize $\Delta \beta$} \\
    \hline

$1.4 M_{\odot}$+$10 M_{\odot}$ \quad & ET+DECIGO  \quad & $1.15 \times 10^{-5}$ \quad& $2.15 \times 10^{-3}$ \quad& $9.73 \times 10^{-8}$ \quad& $2.69 \times 10^{-4}$ \quad& $6.62 \times 10^{-2}$\quad\\
$1.4 M_{\odot}$+$10 M_{\odot}$ \quad & ET+eLISA \quad & $4.60 \times 10^{-5}$ \quad& $1.98 \times 10^{-1}$ \quad& $1.07 \times 10^{-4}$ \quad& $4.53 \times 10^{-2}$ \quad& $1.28$\quad\\
$1.4 M_{\odot}$+$10 M_{\odot}$ \quad & aLIGO+DECIGO  \quad & $6.72 \times 10^{-5}$ \quad& $3.70 \times 10^{-3}$ \quad& $1.13 \times 10^{-7}$ \quad& $3.27 \times 10^{-4}$ \quad& $1.3 \times 10^{-1}$\quad\\
$1.4 M_{\odot}$+$10 M_{\odot}$ \quad & aLIGO+eLISA \quad & $8.41 \times 10^{-4}$ \quad& $3.99$ \quad& $3.85 \times 10^{-4}$ \quad& $1.00$ \quad& $2.52 \times 10^{1}$\quad\\
\hline
\hline
$1.4 M_{\odot}$+$100 M_{\odot}$ \quad & ET+DECIGO  \quad & $7.10 \times 10^{-6}$ \quad& $5.65 \times 10^{-3}$ \quad& $1.30 \times 10^{-7}$ \quad& $1.44 \times 10^{-4}$ \quad& $5.53 \times 10^{-2}$\quad\\
$1.4 M_{\odot}$+$100 M_{\odot}$ \quad & ET+eLISA \quad & $1.27 \times 10^{-5}$ \quad& $6.75 \times 10^{-2}$ \quad& $2.67 \times 10^{-4}$ \quad& $6.2 \times 10^{-3}$ \quad& $6.11 \times 10^{-1}$\quad\\
$1.4 M_{\odot}$+$100 M_{\odot}$ \quad & aLIGO+DECIGO  \quad & $4.31 \times 10^{-5}$ \quad& $1.77 \times 10^{-2}$ \quad& $3.87 \times 10^{-7}$ \quad& $3.94 \times 10^{-4}$ \quad& $2.00 \times 10^{-1}$\quad\\
$1.4 M_{\odot}$+$100 M_{\odot}$ \quad & aLIGO+eLISA \quad & $2.03 \times 10^{-4}$ \quad& $1.41$ \quad& $3.58 \times 10^{-4}$ \quad& $1.50 \times 10^{-1}$ \quad& $1.36 \times 10^1$\quad\\
\hline
 \hline
  \end{tabular}
    \label{tab8}
\end{table}


\begin{thebibliography}{99}

\bibitem{will} C. M. Will, {\it The Confrontation between General Relativity and Experiment},
\lrl, {\bf 9} 3 (2006) (cited [6 March 2015]).

\bibitem{RN} R. Narayan, J. E. McClintock, in \emph{"General Relativity and Gravitation: A Centennial Perspective"}, Editors: A. Ashtekar, B. Berger, J. Isenberg and M.A.H. MacCallum, Cambridge University Press (2013).

\bibitem{kent} K. Yagi, {\it Scientific potential of DECIGO pathfinder and testing GR with space-borne gravitational wave interferometer}, ~\ijmpd
~{\bf 22} 1 (2013).

\bibitem{will_yunes} C. M. Will \& N. Yunes, {\it Testing alternative theories of gravity using LISA}, ~\cqg ~{\bf 21} 4367 (2004).

\bibitem{siemens} N. Yunes \& X. Siemens,{\it Gravitational-Wave Tests of General Relativity with Ground-Based Detectors and Pulsar-Timing Arrays},
\lrl ~{\bf 16} 9 (2013), http://relativity.livingreviews.org/Articles/lrr-2013-9/.


\bibitem{grt1} C. M. Will, {\it Testing scalar-tensor gravity with gravitational-wave observations of inspiralling compact binaries},
~\prd {\bf 50} 6058 (1994).

\bibitem{grt2} E. Berti A. Buonanno \& C. M. Will, {\it Estimating spinning binary parameters and testing alternative theories of gravity with LISA},
\prd {bf 71} 084025 (2005).

\bibitem{grt3} C. K. Mishra et al., {\it Parametrized tests of post-Newtonian theory using Advanced LIGO and Einstein Telescope}
~\prd {82} 064010 (2010).

\bibitem{grt4} K. G. Arun et al., {\it Probing the nonlinear structure of general relativity with black hole binaries}, ~\prd {\bf 74} 024006 (2006).

\bibitem{ppn2} N. Yunes \& F. Pretorius, {\it Fundamental theoretical bias in gravitational 
wave astrophysics 
and the parametrized post-Einsteinian framework}, \prd {\bf 80} 122003 (2009).

\bibitem{ppn1} N. Cornish et al., {\it Gravitational wave tests of general 
relativity with the parameterized post-Einsteinian framework}, \prd
{\bf 84} 062003 (2011).

\bibitem{YT} K. Yagi and T. Tanaka, Phs. Rev. D{\bf 81} 064008 (2010).

\bibitem{luc} L. Blanchet, {\it Gravitational Radiation from Post-Newtonian Sources and Inspiralling Compact Binaries}, \lrl ~{\bf 17} 2 (2014),
http://relativity.livingreviews.org/Articles/lrr-2014-2/.

\bibitem{satya} B. S. Sathyaprakash \& B. F. Schutz, {\it Physics, Astrophysics and Cosmology 
with Gravitational Waves}, \lrl {\bf 12} (2009) (cited [6 March 2015]).

\bibitem{cutler} C. Cutler \& E. E. Flanagan, {\it Gravitational waves from merging compact 
binaries: How accurately can one extract the binary's parameters from the inspiral waveform?},
~\prd {\bf 49} 6 (1994).

\bibitem{veccio} A. Vecchio, Phys. Rev. D {\bf 70} 042001 (2004).

\bibitem{NordWill} K. Nordtvedt, Phys. Rev. {\bf 169} (1968) 1017; C. M. Will and K. Nordtvedt, Astrophys. J {\bf 177} (1972) 757.

\bibitem{moore} C. J. Moore, R. H. Cole and C. P. L. Berry,  Classical \& Qunatum Gravity, {\bf 32} (2015) 015014.


\bibitem{Seto:2001qf} N. Seto , S. Kawamura \& T. Nakamura, 
{\it Possibility of direct measurement of the acceleration of the Universe using 0.1-Hz 
band laser interferometer gravitational wave antenna in space}, ~\prl {\bf 87} 221103 (2001).

\bibitem {seto} K. Yagi \& N. Seto, {\it Detector configuration of DECIGO/BBO and 
identification of cosmological neutron-star binaries}, ~\prd {\bf 83} 044011 (2011).

\bibitem{elisa} P. Amaro-Seoane et al., {\it Low-frequency gravitational-wave science with 
eLISA/NGO}, \cqg {\bf 29} 124016 (2012).

\bibitem{ajith} D. Keppel \& P. Ajith, {\it Constraining the mass of the graviton using 
coalescing black-hole binaries}, \prd {\bf 82} 122001 (2010).

\bibitem{finn} L. S. Finn, {\it Detection, measurement, and gravitational radiation},
~\prd {\bf 46} 12 (1992).

\end{thebibliography}
\end{document}